\documentclass[11pt,twoside]{article}
\usepackage{asp2010}
\usepackage{natbib}
\usepackage{graphicx}

\resetcounters

\begin{document}

\title{Solar like oscillations in the stars KIC~5390438 and KIC~5701829 observed by Kepler}
\author{L. Fox Machado$^1$ and A.S. Baran,$^{2}$
\affil{$^1$Observatorio Astron\'omico Nacional, Instituto de
Astronom\'{\i}a -- Universidad Nacional Aut\'onoma de M\'exico, Ap.
P. 877, Ensenada, BC 22860, M\'exico}
\affil{$^2$Mt. Suhora Observaroty of the Pedagogical University, ul. Podchor\c{a}\.{z}ych 2, 30-084 Cracow, Poland}
}

\begin{abstract}

 The preliminary results of an analysis of the  KIC~5390438  and KIC~5701829  light curves 
are presented.  The variations of these stars were detected  by \citet{baran2011a}  in a
 search for pulsating M dwarfs in the Kepler public database.
The objects have been observed by the Kepler spacecraft during the
Q2 and Q3 runs in a short-candence mode (integration time of $\sim$ 1 min). A Fourier analysis of the time series 
data has been
performed by using the PERIOD04 package. The resulting power spectrum of each star shows a 
clear excess of power in the frequency
range 100 and 350 $\mu$Hz with a sequence of spaced peaks typical of solar-like oscillations. 
A rough estimation of the large and small separations has been obtained. 
  Spectroscopic observations 
secured at the Observatorio Astronomico Nacional in San Pedro M\'artir allowed us to derive a 
spectral classification
 K2III and K0III for KIC~5390438  and KIC~5701829, respectively. 
 Thus, KIC 5390438 and KIC 5701829 have been identified as solar-like oscillating red giant stars.
\end{abstract}
 
\section{Introduction}
\label{introduction}

The Kepler satellite \citep{borucki}, successfully launched in 
2009 March,  is providing  light curves of impressive quality 
  with the primary goal to detect Earth-size planets by means of the transit method. 
 The Kepler spacecraft has onboard  a 0.95-meter diameter telescope with an array of 42 CCDs which
  is observing continuously a fixed field of view of $\sim$ 105 square degrees 
 in the constellations Cygnus and Lyra.  
During the life time of the mission  time series data 
 of more than 150,000 stars will be obtained with a high duty cycle typically from few weeks to several months. 
 Most of these objects will be observed at a near regular long cadence of $\sim$ 30 minutes,
while 512 at a time with a short cadence period of $\sim$ 1-minute.
The high-precision photometry obtained by the satellite is especially well 
suited for probing the interior of the stars by using the techniques of asteroseismology 
\citep{aerts}.  
This database is unique since it can be used to characterize common oscillation properties of 
a large number of intrinsic pulsators by analyzing their amplitude spectra
 \citep{chaplin, hekker, uytter, baran2011a}

In particular, \citet{baran2011a} carried out an analysis of a sample of 86 low mass red stars
aimed at determine whether radial pulsation are excited to detectable amplitudes in M dwarfs as 
suggested recently by theoretical predictions \citep{baran2011b, rodri}.
 \citet{baran2011a} derived the spectral types of these 86 objects after which only six of them
turned out to be main sequence M stars and none out of six showed signals which could be attributed
to M dwarf pulsations. Among these 86 objects a few K-type stars showed short period pulsations
which could be attributed to solar-like oscillations. We present here the first results of an analysis
of two of these stars: KIC~5390438 and KIC~5701829.

\begin{figure}[!ht]
\centering
\includegraphics[height=10cm,width=9.0cm]{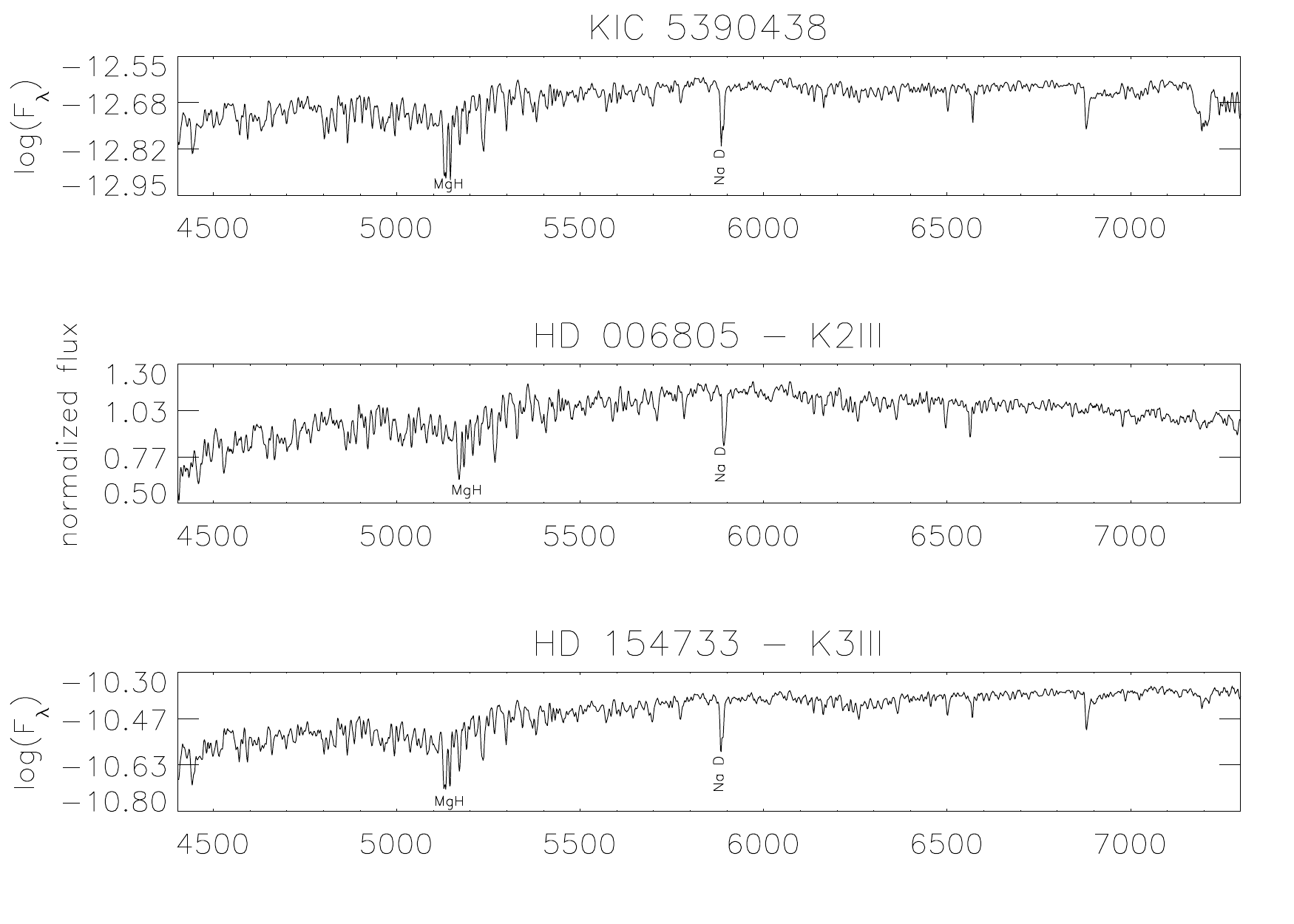}
\caption{A comparison of the spectrum of KIC~5390438 with two MK standards.
From top to bottom: KIC~5390438, K2III standard, K3III standard. The
positions of the Mg I triplet and Na doublets absorption features are indicated in
each panel.}
\label{fig1}
\end{figure}

\begin{figure}[!ht]
\centering
\includegraphics[height=10cm,width=9.0cm]{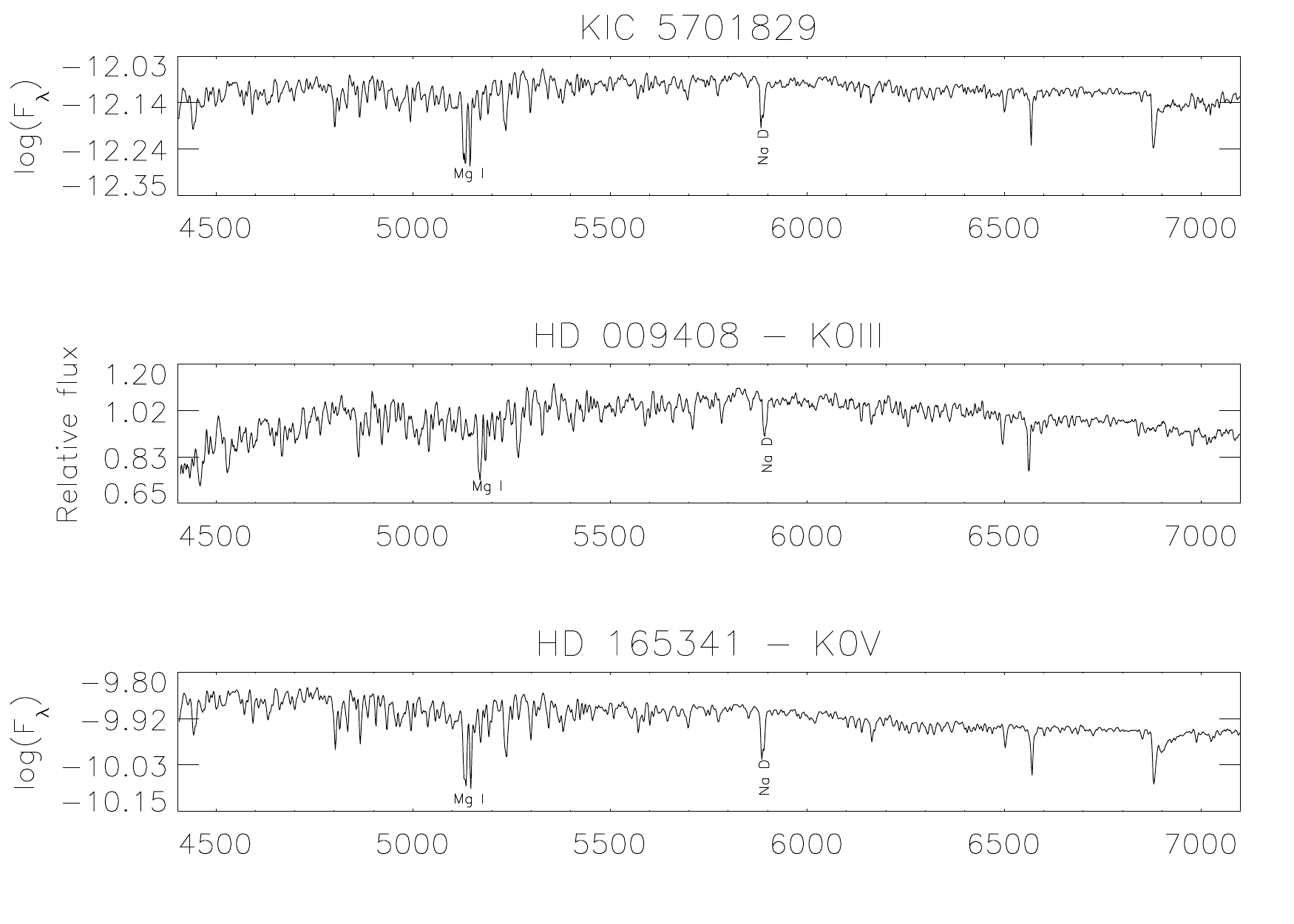}
\caption{A comparison of the spectrum of KIC~5701829 with those of MK standards.
From top to bottom: KIC~5701829, K0III standard, K0V standard. The positions
of the Mg I triplet and Na doublets absorption features are indicated in each
panel.}
\label{fig2}
\end{figure}

\begin{table}[!ht]
\caption{Observational propierties of the target stars as taken from the Kepler
Input Catalogue.\label{table1}}
\begin{center}
 \begin{tabular}{cccccccc}
\hline
KEPLER& RA (2000)& DEC (2000)& Kp Mag& $T_{\rm eff}$& $\log\,g$& E(B-V)& Radius\\
ID &&&(mag)& (K)& (cm/s$^2$)& (mag)&($R_{\odot}$)\\
\hline
5390438 & 19 54 32.22 & +40 35 51.0 &10.538& 4536& 4.418& 0.014& 0.924\\
5701829 &19 22 22.64 &+40 59 42.5 &9.047& 4623 &4.634 &0.006 &0.694\\
\hline
 \end{tabular}
\end{center}
\end{table}

\section{Spectroscopic data}

Details on spectrosopic observations are given in \citet{baran2011a}. Briefly, 
the spectra of KIC~5390438 and KIC~5701829 were acquired at Observatorio Astron\'omico Nacionat at
San Pedro M\'artir in Baja California, Mexico. We used the Italian Boller and Chivens
spectrograph attached to the Cassegrain f/7.5 focus of the 2.12-m telescope. A diffraction
grating with a 400 l/mm ruling and blazed at 6$^{\circ}$ was used in the first order for
the observations, with an inclination angle of 8$^{\circ}$, to cover a wavelength range from
4400 to 7900\,\AA. We employed a dispersion of 1.8\,\AA\, per pixel with a spectral resolution
of 8\,\AA\, (R $\sim$ 750 at central wavelength). For data reduction we followed the standard
procedures using the IRAF package.

The spectral types were defined by direct comparison with template spectra recorded
during the same observing night using the same equipment and with similar resolution
spectra from MILES spectral library \citep{sanchez}. Figure~\ref{fig1} compares
the spectrum of KIC~5390438 (top panel), the spectrum of the K2III
standard HD~006805 (middle panel) and the spectrum of the K3III star HD~154733 (bottom
panel). The second one was obtained from MILES library, while the third one was
acquired with the same equipment. Given that the three spectra are quite similar to each
other, from which we conclude that KIC~5390438 is a K giant star. As a luminosity criterium we used
the Mg I triplet $\lambda \lambda$5167, 5172, 5183, which is sensitive to luminosity for the late-G to
the mid-K. We thus conclude that KIC~5390438 is a K2III star.

Likewise, Figure~\ref{fig2} shows the spectrum of KIC~5701829 (top panel), the
spectrum of the K0 giant star HD~009408 from MILES library (middle panel) and the
spectrum of the K0 main sequence star HD~165341 obtained with the same equipment
(bottom panel). As can be seen the spectrum of the giant star looks similar to its main
sequence counterpart. However, as it is know the features in the main sequence stars are
sharper, in particular the lines of Na I doublets are less stronger in giant stars. Therefore
we conclude that KIC~5701829 is rather a K0III star.

The physical parameters of the stars as taken from the Kepler input catalogue
(KIC) are listed in Table~\ref{table1}. 
Considering the spectral types derived for KIC~5390438  and KIC~5701829,
the values of the effective temperature listed in Table~\ref{table1} are in agreement within the
errors with those of K giant stars, while the values of the surface gravity are overestimated by a
factor of two.  As a reference, the atmospheric parameters
of the standard stars HD~006805 (K2III) and HD~009804 (K0III) are the following:
$T_{\rm eff} = 4600$ K, $\log\, g = 2.90$ cm/s$^{2}$ \citep{hekker1} and 
$T_{\rm eff} = 4804$ K, $\log\, g = 2.30$ cm/s$^{2}$ 
\citep{mishen}, respectively.

\begin{figure}[!ht]
\centering
\includegraphics[height=10cm,width=9.0cm]{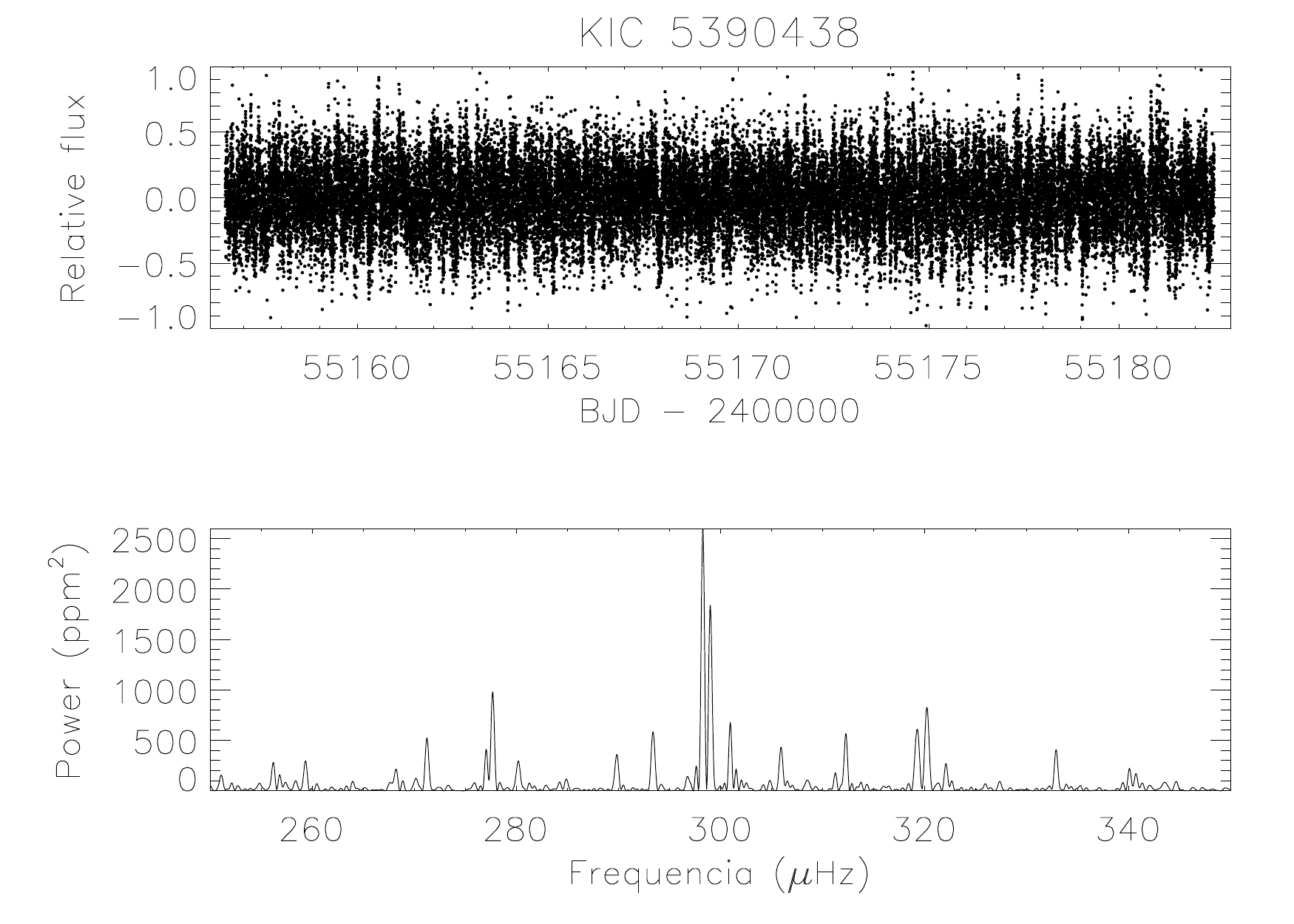}
\caption{Top panel: KIC~5390438 light curve normalized to the mean value.
Bottom panel: Fourier spectrum limited to the range of stellar pulsations.}
\label{fig3}
\end{figure}
\begin{figure}[!ht]
\centering
\includegraphics[height=10cm,width=9.0cm]{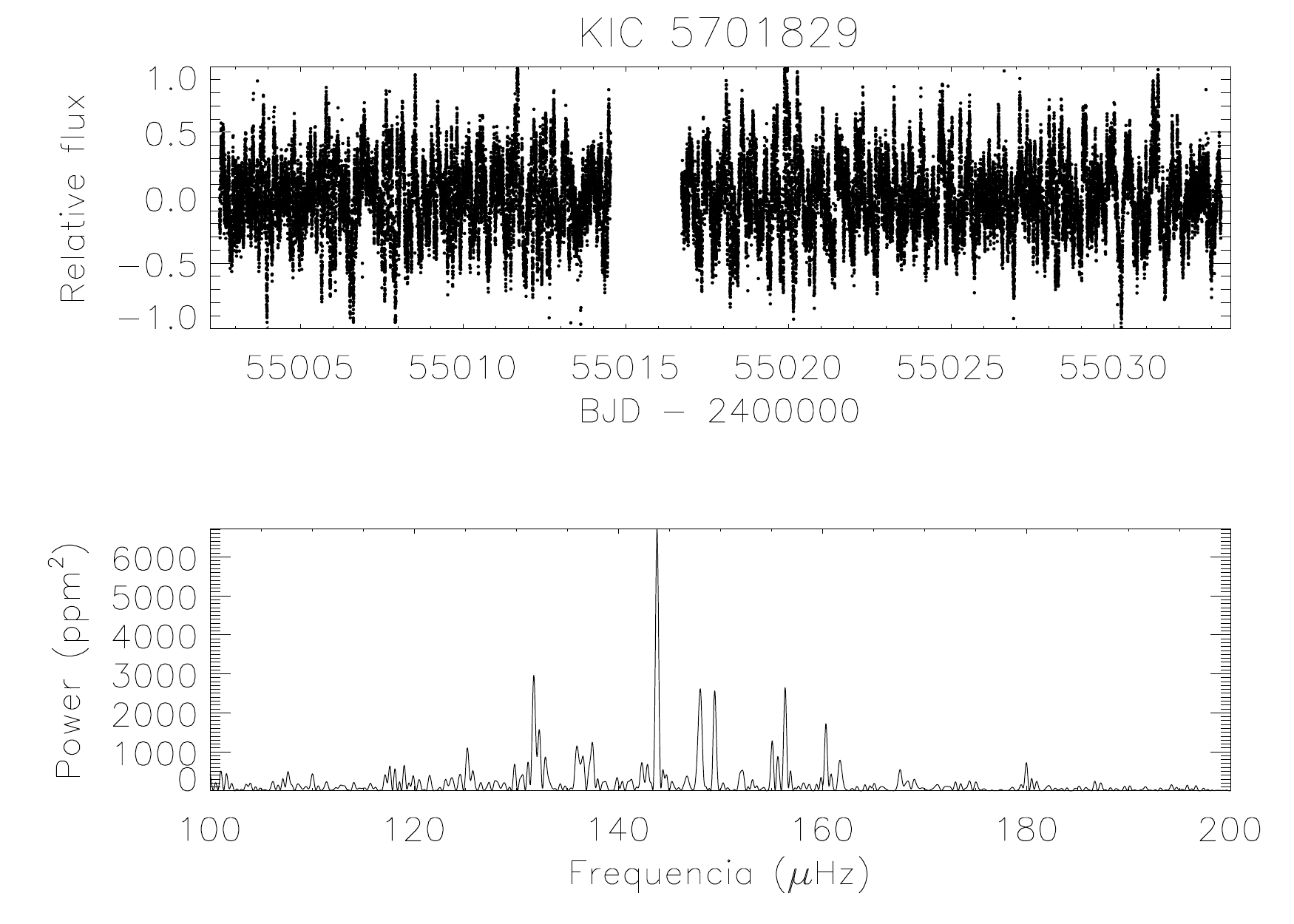}
\caption{Top panel: KIC~5701829 light curve normalized to the mean value.
Bottom panel: Fourier spectrum limited to the range of stellar pulsations.}
\label{fig4}
\end{figure}

\section{Light curves and period analysis}

KIC~5390438 have been observed by the Kepler satellite in a short cadence-mode (integration
time of $\sim$ 1 minute) during 25.98 consecutive days. From \citet{baran2011a}
it follows that it shows intrinsic pulsations with a period of $\sim$ 1 hour. KIC~5701829
data was obtained for a duration of 30.98 days in a short cadence-mode as well. A quick
analysis of the data revealed an intrinsic variation of
about 0.5 hours.

The raw time series data have been processed as explaining in \citet{baran2011a}. 
The light curves
normalized to the mean value are shown in top panels of Fig.~\ref{fig3} and 
Fig.~\ref{fig4} for KIC~5390438 and KIC~5701829, respectively. The Fourier analysis of the data set has been
performed by using the PERIOD04 package \citep{lenz}, which allows to extract the individual
frequencies from large multiperiodic time series and can perform multiple-frequency
least-squares fit of up to several hundred simultaneous sinusoidal components.
The power spectra of KIC~5390438 and KIC~5701829  around the highest peak are shown in bottom
panels of Fig.~\ref{fig3} and Fig.~\ref{fig4}, respectively. We have found that KIC~5390438 shows
a clear power excess in the frequency range 250-350 $\mu$Hz centred at $\nu_{\rm max} \sim 298.3\,\mu$Hz,
while for KIC~5701829 the power excess is located at a lower frequency range between
100-200 $\mu$Hz centred at $\nu_{\rm max} \sim 143.8\, \mu$Hz. 

As well know, the solar-like oscillation spectra are characterized by the large
and small frequency separations. The former consists in a series of equally spaced
peaks separated by $\Delta \nu$ between $p$ modes of the same degree and adjacent $n$: 
$\Delta \nu \sim \Delta \nu_{l}= \nu_{n+1,l} - \nu_{n,l}$.
 The latter is seen as another series of peaks, with separation of
$\delta_{l,l+2} = \nu_{n,l} - \nu_{n-1,l+2}$. While the large frequency separation is proportional to the
square root of the mean density, the small frequency separation is sensitive to the chemical
composition gradient in the central regions of the star and hence to its evolutionary
state. The power spectra presented in Fig.~\ref{fig3} and Fig.~\ref{fig4} allow us to give a rough
estimation of the large and small frequency separations for our target stars.
We have
obtained $\Delta \nu_{0} \sim 21.1\,\mu$Hz, $\delta \nu_{0,2} \sim 0.82\,\mu$Hz and 
$\Delta \nu_{0} \sim 12.2\,\mu$Hz, $\delta \nu_{0,2} \sim 1.46\,\mu$Hz
for KIC~5390438 and KIC~5701829, respectively. We note that the derived values of $\Delta \nu_{0}$ 
are in good agreement with those estimated from the scaling relations for
solar-like oscillations  by \citet{stello}.  

\section{Conclusions}

We have analyzed the light curves of two K-type stars observed by Kepler during aproximately
one month. We provided a spectral classification for each star. A preliminary
analysis of the power spectrum of the light curves allowed us to get a rough estimate
of the large and small frequency separations. 

The empirical scaling relations of \citet{kjeldsen} can be used to provide a 
first estimate of stellar radius and mass without any stellar modelling:

\[ \frac{R}{R_{\odot}} \approx \left (\frac{135\,\mu{\rm  Hz}}{\langle \Delta \nu \rangle}\right)^{2} 
\left( \frac{\nu_{\rm max}}{3050\,\mu{\rm Hz}}\right) \left (\frac{T_{\rm eff}}{5777\,\rm{K}}\right)^{1/2} \]

\[ \frac{M}{M_{\odot}} \approx \left (\frac{135\,\mu{\rm  Hz}}{ \langle \Delta \nu  \rangle}\right)^{4} 
\left (\frac{\nu_{\rm max}}{3050\,\mu{\rm Hz}}\right)^{3} \left(\frac{T_{\rm eff}}{5777\,\rm{K}}\right)^{3/2}\]

For KIC~5390438, we obtain: $\frac{M}{M_{\odot}} \simeq 1.2$ and  $\frac{R}{R_{\odot}} \simeq 4.0$,
while for KIC~5701829, we have $\frac{M}{M_{\odot}} \simeq 1.4$ and  $\frac{R}{R_{\odot}} \simeq 5.8$ .
  The global asteroseismic parameters derived for KIC~5390438 resemble those of the
red giant star KIC~4351319 --a star below the red clump which is
still ascending the red-giant branch \citep{dimauro}.  On the other hand,
KIC~5701829 is slightly more massive star in a more advanced evolutionary state than KIC~5390438. 
These stars will be studied in more details using stellar modelling to retrieve more 
precise stellar parameters.

\acknowledgements LFM acknowledges financial support from the UNAM under grant PAPIIT
IN104612 and from CONACyT by way of grant CC-118611.
Special thanks are given to the technical staff and night assistants of the
San Pedro M\'artir observatory. This project was supported by Polish Ministry of Science
under grant No. N N203 379736. Funding for Kepler Discovery mission is
provided by NASA's Science Mission Directorate. The authors gratefully acknowledge
the entire Kepler team, whose efforts have made these results possible.

\bibliography{lfox}

\end{document}